\documentclass{article}
\usepackage{amsmath,amsthm,amsfonts,amssymb,mathtools,MnSymbol, cite}

\usepackage{caption}
\usepackage{float}
\usepackage{subfig}
\usepackage{enumerate}
\usepackage{url}
\usepackage{todonotes}

\usepackage{geometry}
\begin{document}

\begin{center}
    \LARGE A New Estimate of the Cutoff Value in the Bak-Sneppen Model \\
    \large C. A. Fish\footnote{Fariborz Maseeh Department of Math and Statistics, Portland State University; cafish@pdx.edu}, J. J. P. Veerman\footnote{Fariborz Maseeh Department of Math and Statistics, Portland State University; veerman@pdx.edu}\\
    \small \today
\end{center}

\begin{abstract}
    We present evidence that the Bak-Sneppen model of evolution on $N$ vertices requires $N^3$ iterates to
    reach the stationary state. This is substantially more than previous authors suggested (on the order of $N^2$).
    Based on that estimate, we present a novel algorithm inspired by previous rank-driven analyses
    of the model allowing for direct simulation of the model with populations of up to $N = 25600$ for
    $2\cdot N^3$ iterations. These extensive simulations suggest a cutoff value of $x^* = 0.66747  \pm 0.00025$,
    a value slightly higher than previously estimated.    We also study how the cutoff values $x^*_N$ at finite $N$
    approximate the conjectured value $x^*$ at $N=\infty$. Assuming $x^*_N-x^*_\infty \sim N^{-\nu}$,
    we find values of $\nu$ which are significantly lower than previous estimates.
   
    We also present evidence supporting earlier conjectures that the bulk of the replacements in the Bak-Sneppen model occur in a decreasing fraction
    of the population as $N \to \infty$.

\end{abstract}

\small
\textbf{\textit{Keywords---}} Bak-Sneppen model, critical threshold, rank driven dynamical systems, discrete dynamical systems, self-organized criticality, one-dimensional cellular automata
\normalsize

\section{Introduction}

The Bak-Sneppen model (BS) is a discrete dynamical system defined as follows. Choose a population size
$N \in \mathbb{N}$ and consider a cyclic graph with $N$ nodes. Assign to each node a \textit{fitness}
value uniformly at random from $[0,1] \subset \mathbb{R}$. Name these fitness values $(x_1, x_2, ...
x_N)$ so that $x_N$ and $x_1$ are neighbors (i.e. $x_{N+1} := x_1$ and $x_0 := x_N$). At each time step,
determine the node $x_j$ with smallest fitness (the \textit{weakest}) and replace both it and its
neighbors $x_{j-1}$ and $x_{j+1}$ with fresh values chosen uniformly at random from $[0,1]$ (we say the
minimum node has \textit{mutated}).

Via this process, the weakest nodes are repeatedly eliminated in a way reminiscent of biological
evolution. The mutation of nodes neighboring the weakest coarsely reflects the effects of disappearing
species on nearby species. Bak and Sneppen in \cite{bak} suggest that this model has dynamic features
also seen in biological evolution, namely ``Self-organized criticality'' (see also~\cite{gould} for
discussion of ``punctuated equilibrium'' and~\cite{deboer} for precise exploration of these features).

Observe that if the fitness of the weakest node $x_j$ decreases in a mutation, then the new weakest is
one of $x_j$, $x_{j-1}$ or $x_{j+1}$. In any case $x_j$ will mutate on the next time step. Thus there is
an ``upward pressure'' counteracted by the fact that strong nodes are always at risk of being neighbors
to the weakest. Indeed, it was shown in~\cite{meester} that the mean of the (time) limiting distribution
of fitnesses is bounded away from 1, as the population size $N \to \infty$ (see~\cite{veermanprieto} for
a survey of related results).

After an initial period during which the overall fitness increases, the model reaches a stationay state
with the vast majority of the values above a certain \textit{cutoff value}. Computer simulations suggest
that, for each $N$, the (time) limiting distribution of fitnesses is uniform on $[x_{N}^*, 1]$ for some
cutoff $x_{N}^*$ depending on $N$. Moreover, it appears that as $N \to \infty$, the sequence of cutoff
values $x_{N}^*$ approaches a limiting value close to 2/3. This cutoff value $x^*$ has been
experimentally estimated as $0.66702 \pm 0.00008$~\cite{grassberger}, $0.66702 \pm 0.00003$~\cite{pac},
and $0.6672 \pm 0.0002$~\cite{garcia}. It was shown in~\cite{meester2} that, under certain assumptions,
the cutoff is strictly less than $0.75$.

We found that on the order of $N^3$ time steps are required to reach the stationary state (see Figure \ref{timing}
and Table \ref{cvm}). Since at each timestep the weakest node must be found (which naively takes $O(N)$ time),
the overall naive runtime is $O(N^4)$. Thus it is computationally intensive to estimate the cutoff using naive iteration of the model.
Because the model doesn't lend itself to parallelization, it is difficult to find a quicker way to run
reliable simulations. Indeed, prior literature has focused on finding models which are
less intensive to simulate and give the same exponents and cutoff. In particular, \cite{grassberger} and
\cite{pac} study the ``BS branching process'' in which the system is initialized with conditions close
to the stationary state, and statistical properties of the avalanches are measured. These properties, together with an assumption about the scaling behavior of the avalanches, allow for an estimation of the cutoff. In the interest of supporting these results, we have made use of certain characteristics of the ``rank dynamics'' of the model to efficiently perform a direct simulation, making no assumptions about the scaling behavior. Our simulations of the model are the largest to date: namely $2\cdot N^3$ time steps for $N = 25600$. In comparison, Garcia et al.~\cite{garcia}
simulated $N = 8000$ for only $O(N^2)$ time steps. 

By simulating the model directly, we obtain empirical fitness distributions in which the cutoff can be ``seen'' clearly (see Figure \ref{timing}). Since it is not immediately clear how to best estimate the cutoff from this data, we employed three different methods to obtain a range of possible values. Firstly (``Method 1''), we take the first index at which the value of the histogram
is above a threshold. This method is expected to underestimate the cutoff (due to always taking the leftmost index), and indeed gives
an estimate $x^* = 0.66688 \pm 0.00004$. Secondly (``Method 2''), we measure
the heights of the histograms, making use of the fact that the fitness distribution approaches uniform on $(x^*, 1)$ as
$N\to \infty$, arriving at $x^* = 0.66717 \pm 0.00010$. Finally (``Method 3''), we model the distribution
of fitness values with a density function $\rho_N$ having parameters that scale with $N$. This method takes into account
information about the entire shape of the fitness distribution (rather than a single measurement, as in our other two methods).
Via this method, we estimated a value of $x^* = 0.66747 \pm 0.00025$, a value
slightly (but significantly) higher than previously estimated. 

In addition, we studied how the cutoff values $x^*_N$ at finite $N$ approximate the conjectured value $x^*$ at $N=\infty$.
Assuming $x^*_N-x^*_\infty \sim N^{-\nu}$, we find that $\nu=0.717\pm 0.027$ (via Method 3),
$\nu = 0.728 \pm 0.017$ (via Method 2), or $\nu = 0.978 \pm 0.024$ (via Method 1), all of which are significantly lower than previous estimates
(e.g. \cite{garcia} gives $\nu\approx 1.4$).

\begin{figure}[]
    \centerline{\includegraphics[width=360px]{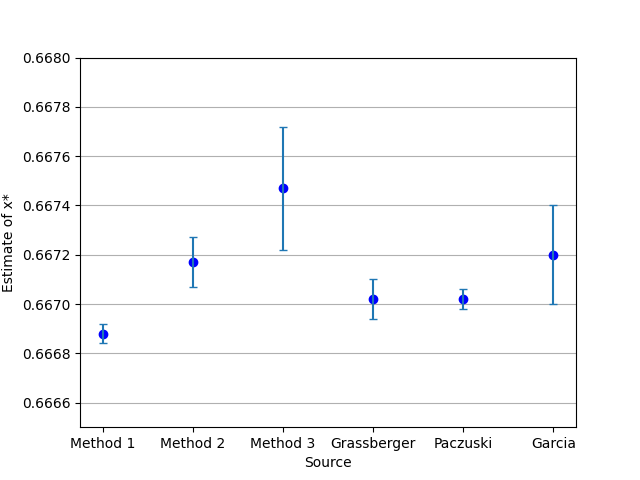}}
    \caption{Comparison of our estimated cutoff values and those estimated by Grassberger~\cite{grassberger}, Paczuski~\cite{pac}, and Garcia~\cite{garcia}. As detailed in Section 4, we used three separate methods to obtain the cutoff values shown.}
    \label{cutoffs}
\end{figure}

\section{Dynamical Characteristics of the Model}

\subsection{\textit{Onset of the stationary state}}

\begin{figure}[]
\centering
    \begin{tabular}{cc}
    \subfloat[$p = 2.4$]{\includegraphics[width = 2.6in]{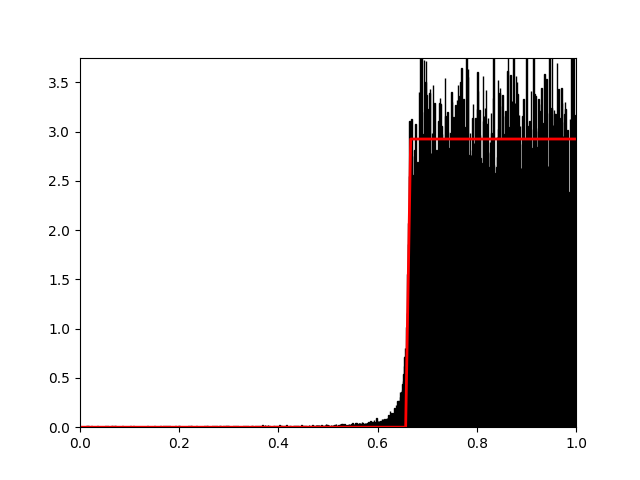}} &
    \subfloat[$p = 2.6$]{\includegraphics[width = 2.6in]{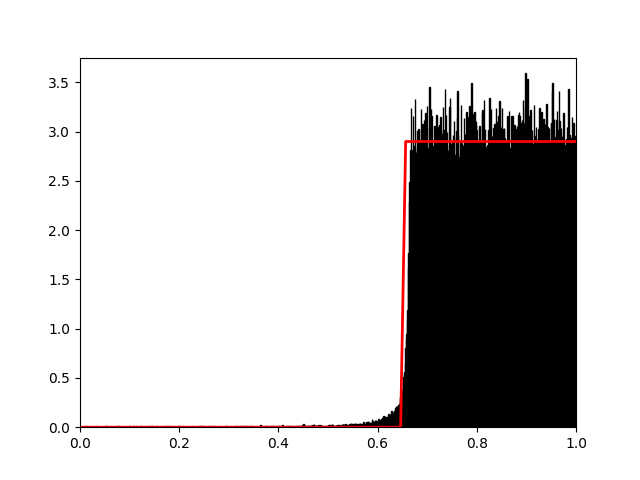}} \\
    \subfloat[$p = 2.8$]{\includegraphics[width = 2.6in]{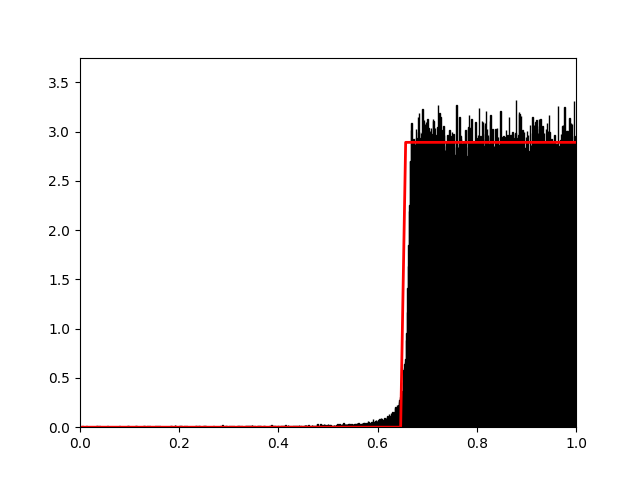}} &
    \subfloat[$p = 3.0$]{\includegraphics[width = 2.6in]{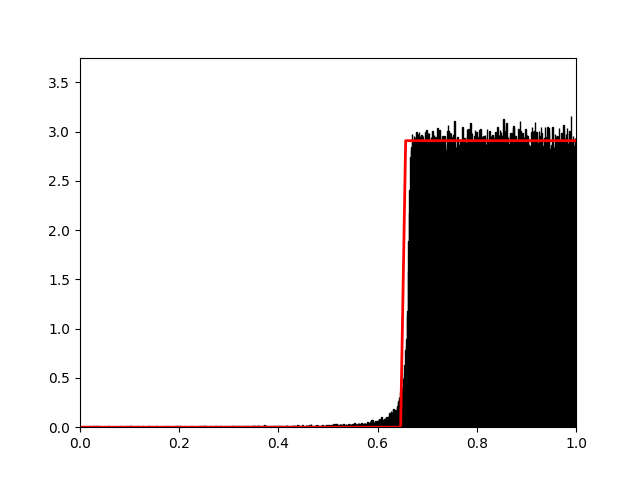}} \\
    \subfloat[$p = 3.2$]{\includegraphics[width = 2.6in]{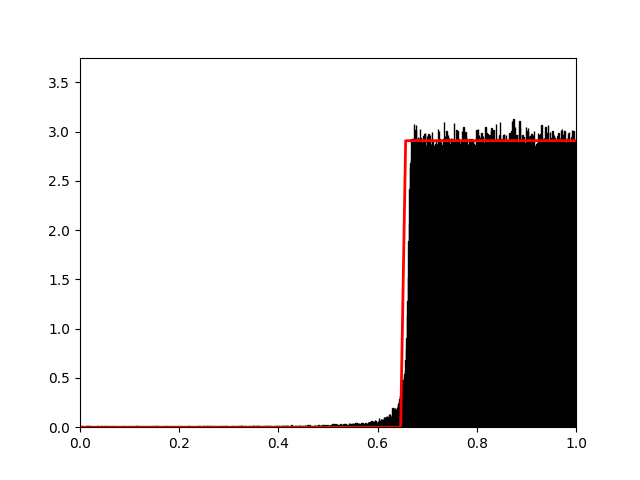}} &
    \subfloat[$p = 3.4$]{\includegraphics[width = 2.6in]{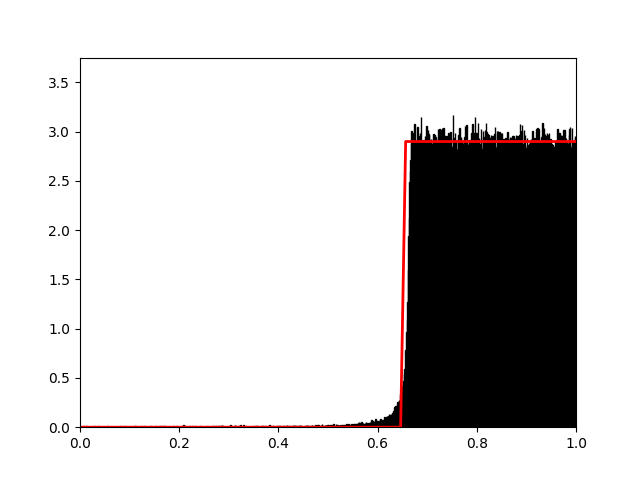}} \\
    \end{tabular}
    \caption{Histograms of fitness values for $500$ population samples made after $N^p$ time steps,
    for $N=800$ and for $p \in \{2.4,2.6,2.8,3.0,3.2,3.4\}$. These data suggest that the stationary state is reached
    at $p \approx 3$. The uniform distribution on $(x^*, 1)$ has been overlaid, for $x^*$ determined individually for each
    histogram (by Method 2, described below). In Table \ref{cvm} we report the Cramér-von Mises and Kolmogorov-Smirnov
    statistics for these data.\label{timing}}

    \begin{tabular}{cccc}
&&&\\
    p & $x^*$ & C.V.M. statistic & Kolmogorov-Smirnov statistic\\ \cline{1-4}
    $2.4$ & 0.658 & 8.325 & 0.0186   \\
    $2.6$ & 0.655 &  1.293 & 0.0160    \\
    $2.8$ & 0.654 & 0.915 & 0.0154      \\
    $3.0$ & 0.656 & 1.043 & 0.0164  \\
    $3.2$ & 0.656 & 1.001 & 0.0162  \\
    $3.4$ & 0.655 & 0.855 & 0.0155

    \end{tabular}
    \captionof{table}{We computed the Cramér-von Mises and Kolmogorov-Smirnov statistics (see e.g. \cite{cvm_encyclo})
    on the empirical distribution of fitness values
    sampled $500$ times after $N^p$ time steps, for $N = 800$. The statistic gives a measure of distance
    between the empirical distribution of the sample and the uniform distribution on $(x^*,1)$. As seen, the
    CVM statistic is somewhat larger for $p \in \{2.4,2.6\}$ but is more consistent for higher values, suggesting that
    at least $p \approx 2.8$ is needed. \label{cvm}}

\end{figure}

Our approach is to iterate the model $T_N$ times on $N$ nodes until it reaches the stationary state. Once it
does, we iterate it another $T_N$ times during which we perform our measurements. This way
we insure that measurements are taken while the system is in the stationary state. Figure \ref{timing} provides
evidence that the stationary state sets in after approximately $T_N=N^3$ iterates of the system. We found similar
behavior for other values of $N$.

We emphasize that this analysis is only meant to suggest the approximate number of iterations required to reach the
stationary state. We will later demonstrate a more robust method of fitting these fitness distributions to a density function.

\subsection{\textit{Most Replacements Have Low Rank}}

\begin{figure}[]
    \centerline{\includegraphics[width=320px]{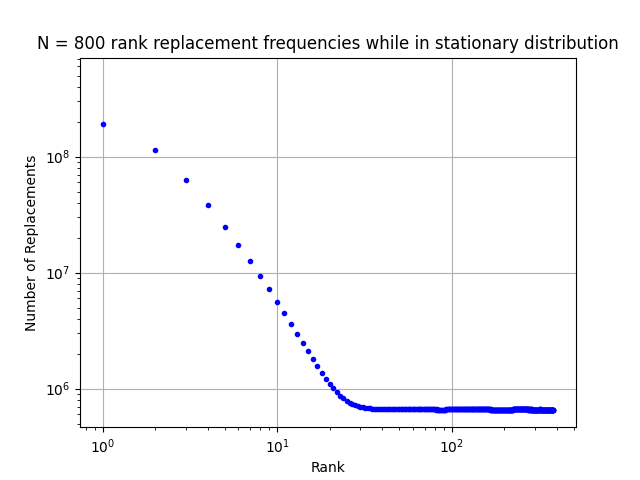}}
    \caption{The frequency of mutation by rank found in one execution of the Bak-Sneppen model at
    $N = 800$. The minimum node (rank 0) is replaced the most often (since it is mutated every time
    step), and it appears that all but the lowest-ranked nodes are mutated rarely and at roughly equal
    frequency. (Note the logarithmic axes.)}
    \label{rankreplacement}
\end{figure}

We can order the values of the fitness distribution by their \emph{rank}. The lowest fitness is ranked ``1",
the next lowest gets rank ``2", et cetera.
The rank-driven analysis put forward in \cite{grinfeld}, \cite{grinfeld2}, and \cite{veermanprieto}
strongly suggests that the vast majority of mutations occur at the lowest ranks.
For example, Figure \ref{rankreplacement} illustrates the frequency at which
the different ranks are replaced (after reaching the stationary distribution) for one simulation at
$N =800$. The overwhelming majority of replacements occur below rank 50. This motivates maintaining a
``short list'' containing
only the lowest-ranked nodes. Mutated nodes may enter and leave the short list, where they stay ``waiting''
to be mutated. If the list grows too small, it is rebuilt by sorting the entire population.
Keeping track only of relatively few lowest-ranked nodes on the short list allows us to find
the position of the lowest ranked node without searching the whole population. The trick is
to keep the list long enough so that we avoid sorting the entire population too often. This strategy
allows us to iterate the model $2\cdot N^3$ times in approximately $O(N^3 \log N)$ time (compare the naive
method at $O(N^4)$ and a standard heap method at $O(N^3 \log N)$ as discussed in Section 3.1).

\subsection{\textit{High Rank Replaced at Constant Rate}}

The rank-driven approach of \cite{veermanprieto} implies that the cutoff can be precisely determined if
the sequence (indexed by the population size $N$) of sequences (indexed by rank $i$) of rank replacement frequencies $\{\alpha_i(N)\}_{i=1}^N$
form a so-called $(p,q)$-array. This is an array $\{\alpha_i(N)\}_{i=1}^N$ such that there exist
$p$, $q$ independent of $N$ with
\[
\sum_{i=0}^{K(N)} \alpha_i \leq p < \sum_{i=1}^{K(N)+1} \alpha_i \quad \textrm{and} \quad
q = \sum_{i=1}^N \alpha_i \,.
\]
for some $K(N) \in (1,N)$ depending on $N$ such that $K(N) = o(N / \ln N)$. Furthermore, there must exist $\delta_i(N)$ and $\tilde{\alpha}(N)$ depending on $N$ such that for ranks $i > K(N)$, the rank replacement
frequencies $\alpha_i$ take the form $\tilde{\alpha}(N) + \delta_i(N)$ where
$\max_{i>K(N)} \vert \delta_i(N) \vert = o(1/N)$ and $\tilde{\alpha}(N)$ is a constant -- that is, the
tail of the sequence of $\alpha_i(N)$ should increasingly flat as $N$ tends to infinity. If the
replacements frequencies $\{\alpha_i(N)\}$ form a $(p,q)$ array, then one can prove \cite{veermanprieto}
that the cutoff equals $\frac{p}{q}$.

\begin{figure}[]
    \centerline{\includegraphics[width=320px]{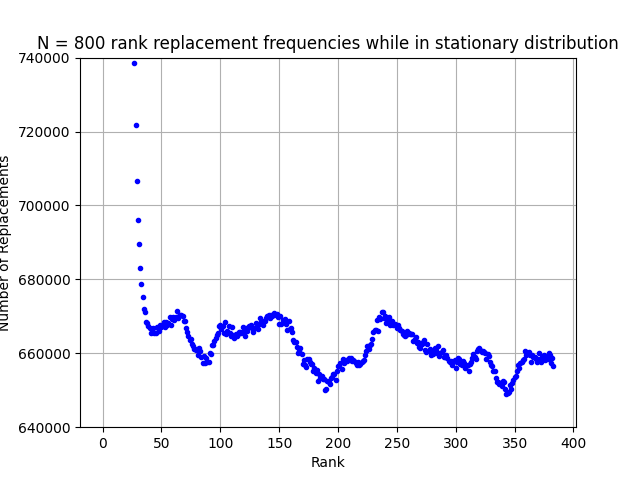}}
    \caption{The frequency of mutation by rank found in one execution of the Bak-Sneppen model at
    $N = 800$, displaying only the vertical range $[640000,740000]$. The data are very close to flat: the
    frequency beyond rank 50 deviates from flat by less than 2 out of 66 or $\pm 1.5\%$. Observe also
    that replacement rates appear to be continuous, perhaps even smooth (rather than random deviations
    from a mean).}
    \label{rankreplacementzoom}
\end{figure}

Our data are consistent with the hypothesis that the sequences $\{\alpha_i(N)\}_{i=1}^N$ form a
$(p,q)$-array, as illustrated in Figure \ref{rankreplacementzoom}. A more rigorous check is difficult in
practice since the precise criterion for selecting $K(N)$ from data alone is unclear, and small errors
in that choice amount to large differences in the estimate of $x^*$. Further investigation would
indeed be useful.

\subsection{\textit{Continuity of High Rank Replacement Rate}}

We also observe an unexpected continuity in the rank replacement frequencies (as in Figure
\ref{rankreplacementzoom}); rank neighbors have correlated replacement frequencies. We propose the
following heuristic explanation, making use of the fact that the distribution, $C(x)$, of the distance $x$
between subsequent mutations is known (e.g. \cite{bak}) to follow a power law $C(x) = x^{-3.15\pm0.05}$.
Let $R_t(x_i)$ be the rank of node $x_i$ at time $t$, and let $P_t(r)$ be the probability that the node with
rank $r$ is mutated at time $t$. Suppose that at time $t$ the least-fit node is $x_m$, and so will mutate next.
Consider some node $x_k$, distinct from $x_m$, $x_{m-1}$, and $x_{m+1}$. Prior to mutation, there is an
approximate chance $p = C(d(x_m, x_k))$ that $x_k$ will be mutated next\footnote{The value $d(x_m, x_k)$
being the minimal distance between $x_m$ and $x_k$, i.e. the minimum of $\vert m - k \vert$ and $N - \vert m - k \vert$.}.
 Since $x_k$ has rank $R_t(x_k)$,
we can say $P_t(R_t(x_k))$ is also approximately $p$. After the mutation, at time $t+1$, the new minimum
is likely\footnote{The new minumum must be one of these three nodes unless all three increase in fitness
as a result of the mutation.} to be one of $x_m$, $x_{m-1}$, or $x_{m+1}$. If so, the chance that $x_k$
is mutated next will be close to $p$, by the continuity of $C(x)$. Hence $P_{t+1}(R_{t+1}(x_k))$ is also
close to $p$. However, $R_{t+1}(x_k)$ may differ from $R_t(x_k)$ due to the mutation, by at most 3.
Thus, for $r = R_t(x_k) \pm 3$, we have $P_{t+1}(r) \approx P_{t}(R_t(x_k)) \approx p$, leading to an apparent
approximate continuity in the graph of rank replacement frequencies.

\section{Simulation Method}

To simulate the model with our ``short list'' method, we used a multi-linked-list data structure to
keep track of the fitnesses, their \textit{position neighbors}, and possibly their \textit{rank
neighbors}. That is, to represent the fitness value $x_j$, we create a node containing:
\begin{itemize}
    \item a \texttt{double}\footnote{a 64-bit floating-point type; see~\cite{kahan} for details}
        approximating $x_j$
    \item a pointer to the node with fitness $x_{j-1}$
    \item a pointer to the node with fitness $x_{j+1}$
    \item (possibly) a pointer to the node with next largest fitness
    \item (possibly) a pointer to the node with next smallest fitness
\end{itemize}
Choosing some $k$ depending on $N$, the simulation is initialized with only the $5k$ lowest-ranked nodes
having pointers to their rank-neighbors. We also maintain pointers to the lowest and highest values on
the short list.

To update the simulation once, the fitness of the lowest-ranked node $x_j$ is replaced, along with the
fitnesses of its two position neighbors $x_{j-1}$ and $x_{j+1}$. It is then determined, for each of the
three nodes $x_j, x_{j-1}, x_{j+1}$, whether they should or should not be on the short list. This is
easy to check by comparing a new fitness value with the fitness of the highest-ranked node on the short
list: the new fitness should be on the short list only if it is strictly less than this highest-ranked
value. The neighbors $x_{j-1}$ and $x_{j+1}$ may or may not be on the short list already, and so may
remain on, remain off, be inserted, or be removed. The rankings of all members of the short list are
maintained appropriately as these operations are performed. If the length of the short list ever
increases above $5k$, the ranks above $5k$ are simply ``forgotten". If it decreases to below $3k$, it
is rebuilt by sorting the population. This last operation is expensive, but occurs infrequently (for
example, in one simulation of 800 nodes lasting $800^3 = 5.12\times 10^8$ iterations, the short list
was rebuilt only $1419$ times; see also Figure \ref{Rgraph}).

This method allows for simulation of large population values in a reasonable time. Our longest
simulations ($N=25600$) required approximately $45$ days to complete\footnote{on the Coeus HPC
Cluster provided through the Portland Institute for Computational Science and Portland State
University}, while carrying out the same simulation using the naive method would take an estimated
several thousand days.

Both the speed of the simulation and the frequency at which the short list is rebuilt will depend
on the choice of $k$. We chose $k = 8$ for $N \in \{100,200,400\}$, $k = 16$ for $N \in
\{800,1600,3200\}$,
$k = 24$ for $N \in \{6400,12800\}$, and $k = 32$ for $N = 25600$. These choices are likely not optimal
and merit further investigation.

\subsection{\textit{Runtime analysis}}

At a given point in time, let $x_{min}$ denote the minumum node and $x_{smax}$ denote the largest node on the short list. In a single interation of the algorithm, the following steps are performed:
\begin{center}
    \begin{enumerate}[(1)]
        \item Determine $x_{min}$ and its position neighbors. Then generate three new random values. For each new value, determine if it is less than $x_{smax}$
        \item If a new value is less than $x_{smax}$, perform a walk through the shortlist to determine its new rank.
        \item Determine if the shortlist needs to be rebuilt by checking if its length is below $3k$. If so, then sort the population and repopulate the shortlist.
    \end{enumerate}
\end{center}
Step (1) is $O(1)$ since pointers to $x_{min}$ and $x_{smax}$ are maintained.

Step (2) is $O(k)$ if it occurs, since the maximum length of the shortlist is $5k$. The probability that a new value is less than $x_{smax}$ is just $x_{smax} < 1$, because it is chosen uniformly at random in $(0,1)$. One may attempt to determine the average value of $x_{smax}$, but since this average is no more than $1$, step (2) is $O(k)$ regardless.

Step (3) is $O(k + N \log N)$ if it occurs, since the population is first sorted with a standard efficient algorithm, and then the shortlist of length $O(k)$ is repopulated. Let $R$ be the probability that this occurs.

\begin{figure}[]
    \centering
    \begin{tabular}{c|c|c}
        N	& Mean number of occurrences &	Estimated $R$\\
        \cline{1-3}	
        100	&54.200	&$2.710 \cdot 10^{-5}$\\
        200	&79.600	&$4.975\cdot 10^{-6}$\\
        400	&325.375&	$2.542\cdot 10^{-6}$\\
        800	&1423.175	&$1.390\cdot 10^{-6}$ \\
        1600&	7073.000&	$8.634\cdot 10^{-7} $\\
        3200&	41310.925&	$6.304\cdot 10^{-7}$\\
        6400&	276566.175&$	5.275\cdot 10^{-7}$\\
        12800&	2023577.800&$	4.825\cdot 10^{-7}$\\
	25600& 2267531.579&$ 6.758\cdot 10^{-8}$\\
    \end{tabular}
    \captionof{table}{For each population value, the number of times the shortlist was rebuilt was recorded and averaged over all runs. Dividing by the total iterations $2N^3$ gives an estimate for $R$, the probability that the shortlist is rebuilt.\label{estimateR}}

    \centerline{\includegraphics[width=300px]{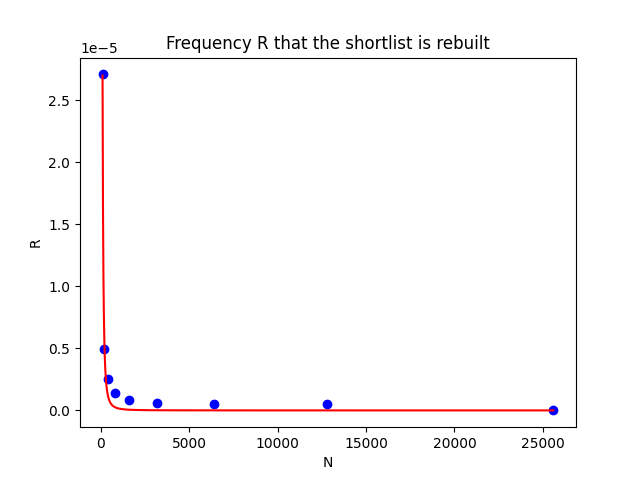}}
    \captionof{figure}{Empirical estimate of $R$, the probability that the shortlist is rebuilt on a given iteration of the algorithm. The fitted curve is given
by $R(N) = 0.805 N^{-2.24}$, giving a growth rate of approximately $O(N^{-2})$.    \label{Rgraph}}

    \centerline{\includegraphics[width=300px]{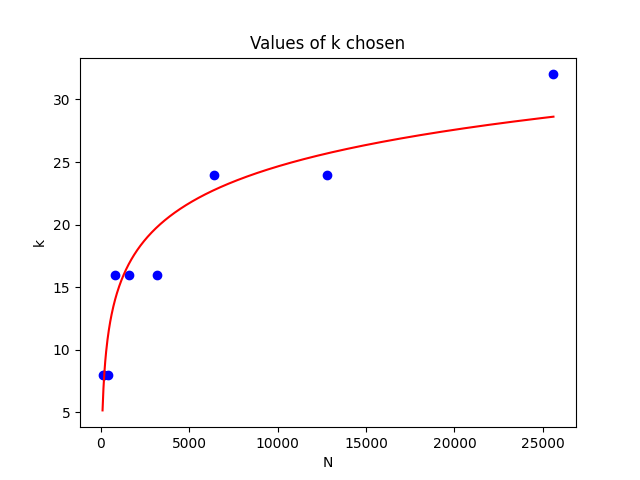}}
    \captionof{figure}{The values of $k$ chosen for our simulations grow as approximately $O(\log N)$. The fitted curve shown is $k(N) = 9.74 \log_{10}(N) - 14.33$. The optimal values of $k$ merit further investigation.    \label{kplot}}

\end{figure}

Thus, a single iteration of the algorithm is on average $O(k + R(k + N \log N))$. Although a theoretical value of $R$ is difficult to determine, an empirical estimate can be found by tracking how often the shortlist is rebuilt (see Table \ref{estimateR}).

If we assume that the probability $R$ grows as  $O(N^{-2})$ (see Figure \ref{Rgraph}), then a single iteration of the algorithm becomes $O(k + N^{-2}k + N^{-1}\log N)$. In our simulations, the value of $k$ was chosen depending on $N$ (see Figure \ref{kplot}). With these choices, $k$ grows roughly as $O(\log N)$. If we assume $k = O(\log N)$, then our average runtime for a single step becomes $O(\log N + N^{-2}\log N +   N^{-1}\log N)$ i.e. $O(\log N)$. This is comparable with a standard heap-based algorithm (e.g.~\cite{grassberger}) which takes $O(\log N)$ time to perform a single iteration. This runtime may be futher reduced by more accurately selecting $k$.

\section{Estimation of the Cutoff}

\subsection{\textit{The cutoff for finite $N$}}

To estimate the cutoff value $x^*_N$ for finite $N$, we performed simulations for various
population values, using two separate pseudorandom number generators (RNGs) in order to mitigate bias:
\texttt{xoshiro256+} (``Xoshiro'', see~\cite{xoshiro}) and \texttt{MT19937-64} (``Mersenne twister'',
see~\cite{mersenne}). For each RNG we performed $20$ simultaneous executions of the model at each
of the population levels $N = 100\cdot 2^i$ for $i \in \{0,...,8\}$.

In each execution, we first iterate $N^3$ times to reach the stationary state. Then, iterate $N^3$ times during which 500 snapshots are taken:
one after each $N^3/500$ iterates. Collect these data into a histogram, in which the cutoff value is
distinctly visible (see Figure \ref{histograms}). Using these histograms, we estimate the cutoff $x^*$ in three different ways.
The raw cutoff data for Methods 1 and 2 is summarized in Table \ref{histogramdata}. We also report ``combined'' values by pooling data
from both RNGs before taking the mean and standard deviation.

\subsubsection{Method 1}

Since we expect the limiting fitness distribution
\[
f(x) = \begin{cases} 0 & 0 \leq x < x^* \\ \frac{1}{1-x^*} & x^* \leq x \leq 1 \end{cases} \,,
\]
with $x^*\approx 2/3$, it is reasonable to define the transition point as the first index in the histogram
at which the height exceeds a threshold $T = 1.5$.

\begin{figure}[]
    \centering
    \begin{tabular}{cc}
    \subfloat[Mersenne: $N = 100$]{\includegraphics[width = 2.5in]{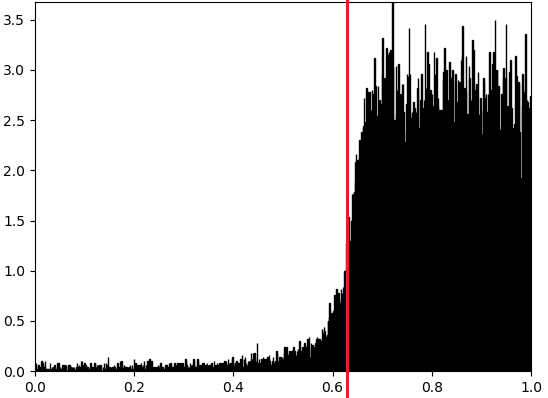}} &
    \subfloat[Xoshiro: $N = 100$]{\includegraphics[width = 2.5in]{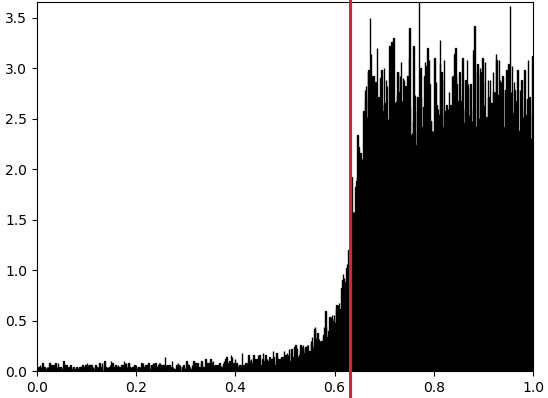}} \\
    \subfloat[Mersenne: $N = 800$]{\includegraphics[width = 2.5in]{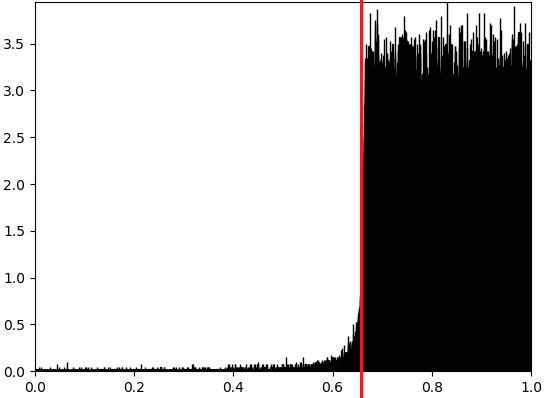}} &
    \subfloat[Xoshiro: $N = 800$]{\includegraphics[width = 2.5in]{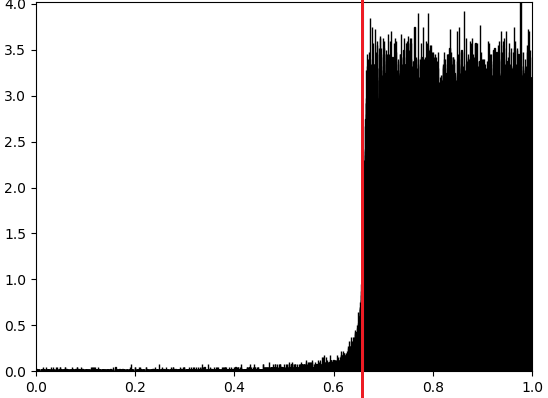}} \\
    \subfloat[Mersenne: $N = 6400$]{\includegraphics[width = 2.5in]{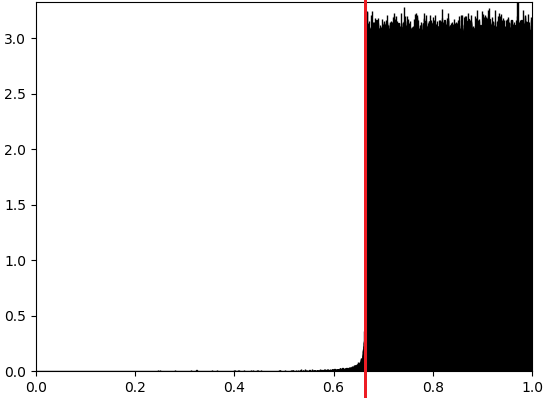}} &
    \subfloat[Xoshiro: $N = 6400$]{\includegraphics[width = 2.5in]{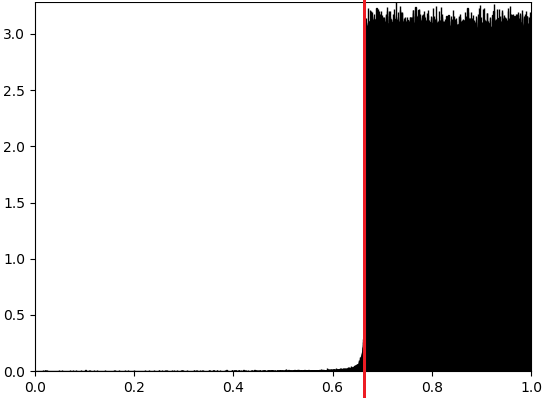}}
    \end{tabular}
    \caption{Histograms of fitness values for $500$ population samples made after reaching the stationary
    distribution. The cutoff is distinct at the higher population counts, and is represented in the figure by
    a red line. Each histogram depicts one of $40$
    executions ($20$ per RNG). They are normalized to have area $1$. }
    \label{histograms}
    \end{figure}

\setlength{\tabcolsep}{.125cm}

    \begin{table}[]
        \begin{center}
    \begin{tabular}{lllllllllllllll}
    i & 0 & 1     & 2     & 3     & 4    & 5    & 6    & 7    & 8    \\
    N & 100 & 200     & 400     & 800     & 1600    & 3200    & 6400    & 12800    & 25600    \\ \cline{1-10} \\
    \textbf{Method 1} &         &         &         &         &         &         &         &          &          \\
    \underline{Mers.} &         &         &         &         &         &         &         &          &          \\
    $\overline{x}^*_i$  & 0.60653 & 0.63885 & 0.65309 & 0.66033 & 0.66347 & 0.66512 & 0.66598 & 0.66637 & 0.66668 \\
    $s_i$ & 0.00916 & 0.00265 & 0.00144 & 0.00074 & 0.00025 & 0.00013 & 0.00005 & 0.00004 & 0.00003 \\
    &         &         &         &         &         &         &         &          &          \\
    \underline{Xosh.} &         &         &         &         &         &         &         &          &          \\
    $\overline{x}^*_i$  & 0.60193 & 0.63981 & 0.65362 & 0.66031 & 0.66349 & 0.66502 & 0.66596 & 0.66635 & 0.66669 \\
    $s_i$  & 0.01197 & 0.00332 & 0.00128 & 0.00033 & 0.00024 & 0.00013 & 0.00006 & 0.00002 & 0.00002 \\
    &         &         &         &         &         &         &         &          &          \\
    \underline{Comb.} &         &         &         &         &         &         &         &          &          \\
    $\overline{x}^*_i$  & 0.60423 & 0.63933 & 0.65335 & 0.66032 & 0.66346 & 0.66506 & 0.66597 & 0.66641 & 0.66668 \\
    $s_i$  & 0.01091 & 0.00304 & 0.00139 & 0.00057 & 0.00025 & 0.00014 & 0.00006 & 0.00003 & 0.00003\\

	 &&&&&&&&&&\\
   \textbf{Method 2} &         &         &         &         &         &         &         &          &          \\
    \underline{Mers.} &         &         &         &         &         &         &         &          &          \\
    $\overline{x}^*_i$  & 0.61211 &0.63368	&0.64661&	0.65509&	0.65967&	0.66274&	0.66446&	0.66557&	 0.66617\\
    $s_i$ & 0.00305	&0.00154&	0.00089&	0.00036&	0.00036&	0.00015	&0.00016	&0.00010&	0.00012 \\
    &         &         &         &         &         &         &         &          &          \\
    \underline{Xosh.} &         &         &         &         &         &         &         &          &          \\
    $\overline{x}^*_i$  & 0.61252&	0.63400&	0.64697&	0.65514	&0.65977	&0.66278&	0.66444&	0.66555&	 0.66618 \\
    $s_i$  &0.00288&	0.00158	&0.00065&	0.00050&	0.00032	&0.00024	&0.00016&	0.00005	&0.00009 \\
    &         &         &         &         &         &         &         &          &          \\
    \underline{Comb.} &         &         &         &         &         &         &         &          &          \\
    $\overline{x}^*_i$  & 0.61231	&0.63384&	0.64679&	0.65511	&0.65972&	0.66276&	0.66445&	0.66556	 &0.66618 \\
    $s_i$  & 0.00297	&0.00157&	0.00080&	0.00044&	0.00035	&0.00020&	0.00016	&0.00008	&0.00011
    \end{tabular}
    \caption{The mean and standard deviation (over $20$ executions) of the cutoff estimated via
    Methods 1 and 2, for each population value and for each RNG (Mersenne twister, Xoshiro) and for the
    combined data (found by pooling data from each RNG before taking the mean and standard deviation).
    Note that the population values are of the form $N = 100\cdot 2^i$ for $i \in \{0,1,2,3,4,5,6,7,8\}$.
    Note also that there values do not correspond directly with Figure \ref{cutoffs}, as the latter values
    are obtained via the randomized data method described below.}
    \label{histogramdata}
\end{center}
    \end{table}

\begin{figure}[]
    \centerline{\includegraphics[width=300px]{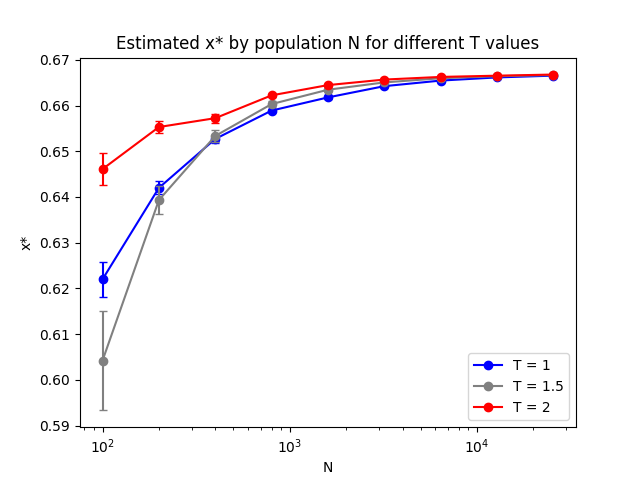}}
    \caption{\textbf{Method 1} estimates the cutoff by determining the first index at which the histogram exceeds a chosen value $T$. Because the limiting
distribution is singular at $x^*$, any choice of $T$ will give the same cutoff value in the limit. Estimated values of the cutoff $x^*$ are plotted
for different populations $N$ and a range of choices of $T$. (Note the logarithmic scale of the $x$-axis.)}
    \label{cutoff_by_T}
\end{figure}

Since we we expect a discontinuity at $x^*$ in the limiting distribution, any choice of $T \in (0,1)$ should (eventually) provide
an identical cutoff value, and the choice of $1.5$ is for convenience. As  Figure \ref{cutoff_by_T} illustrates, the estimate of $x^*$ is
affected by choice of $T$, but the estimates appear to converge in the limit.  The set estimates of $x^*$ via this method has the smallest
standard deviation, but is not ideal since it is is sensitive to noisy data. In particular, it selects the \textit{first} index at which the histogram
exceeds $T$, which does not always capture the ``obvious'' position of the cutoff as histogram is generally not monotonic, and in fact
generally underestimates the cutoff value. This is especially the case at lower population values, where the data is noisy and the approach
to the cutoff is gradual (see the histograms for $N=100$ in Figure \ref{histograms}). This method also does not take into account the overall
shape of the distribution except that it approaches uniform. Methods 2 and 3 attempt to address these issues, but lead to a wider range of values for $x^*$.

\subsubsection{Method 2}

Since the integral of the limiting uniform fitness distribution is $1$, we can estimate $x^*$ by estimating the height $h$ of the distribution,
and then computing $x^* = 1 - h^{-1}$. Moreover, since the limiting height is uniform on $(x^*, 1)$, we can choose any subinterval  of
$(x^*, 1)$ on which to estimate $h$; in particular we can choose an interval with an upper bound of $x^*$ as its left endpoint.
To avoid the noisy ``left edge'' of the rectangle, we chose the interval $(0.8, 1)$ and estimate $h$ by simply averaging the height of the
histogram on that interval. Observe that this method ignores the fact that a non-zero portion of the distribution lies below the cutoff
in every finite empirical fitness histogram. This means that the height of the right-half of the empirical histogram is \textit{lower} than
that of a uniform distribution with the same cutoff value, so that the method will tend to underestimate the cutoff. Method 3 below
addresses this concern.

\subsubsection{Method 3}

\begin{figure}[]
\centering

    \begin{tabular}{lllllllllllllll}
	&&&&&&&&&\\
&&&&&&&&&\\
    N & 100 & 200     & 400     & 800     & 1600    & 3200    & 6400    & 12800    & 25600    \\ \cline{1-10}
    &         &         &         &         &         &         &         &          &          \\
    $\overline{\alpha}_N$  & -0.48111&	-0.53420&	-0.59730&	-0.66636&	-0.57371&	-0.61219&	-0.65546	 &-0.75983	&-0.81332\\
    $s_i$ &0.05772&	0.04334	&0.02149	&0.02445&	0.01894&	0.03071&	0.04463&	0.04704	&0.06853\\
    &         &         &         &         &         &         &         &          &          \\
    $\overline{\beta}_N$  &0.04133	&0.02690	&0.01591	&0.00906&	0.00745	&0.00473&	0.00278	&0.00114&	 0.00065\\
    $s_i$  &0.00848&	0.00341&	0.00090	&0.00083	&0.00100	&0.00081&	0.00069&	0.00042&	 0.00027\\
    &         &         &         &         &         &         &         &          &          \\
    $\overline{\gamma}_N$  &0.61264	&0.63471	&0.64756&	0.65566&	0.65990	&0.66303&	0.66473&	0.66590	 &0.66644 \\
    $s_i$  &0.00312&	0.00256&	0.00118&	0.00074&	0.00042&	0.00030&	0.00027&	0.00016&	 0.00018
    \end{tabular}
    \captionof{table}{\textbf{Method 3}. The mean and standard deviation of sets of best-fit parameters $\alpha_N$, $\beta_N$, and $\gamma_N$ for each population level.
The curve given by$\rho_N(x) = \beta_N(\gamma_N - x)^{\alpha_N}$ for $x \in (0,\gamma_N)$ models the density of the fitness distribution below the cutoff $x^* \approx \gamma_N$.
Above the cutoff the density is the constant $h = (1-\gamma_N)^{-1}$. \label{scaling_data}}

    \includegraphics[width=210px]{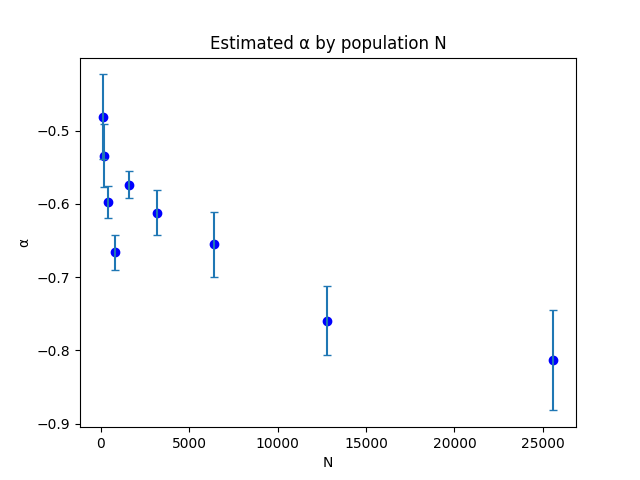}
    \includegraphics[width=210px]{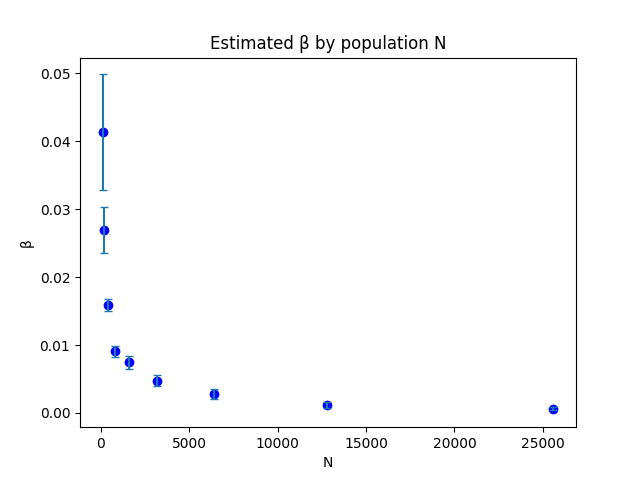}
    \includegraphics[width=210px]{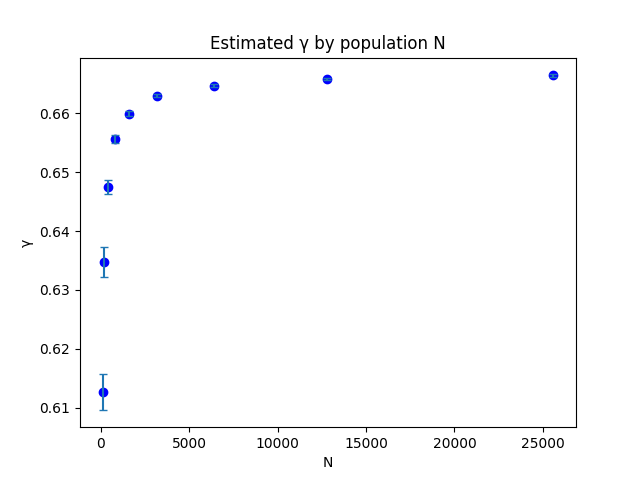}
    \captionof{figure}{\textbf{Method 3.} Estimated parameters $\alpha_N$, $\beta_N$, and $\gamma_N$ for each population level. \label{scaling_parameters}\\}

\end{figure}

One may observe (e.g. Figure \ref{histograms}) that for finite $N$ the fitness distribution has some non-zero density in the interval $(0,x^*)$,
which appears (at all population levels) to roughly follow a power law. This suggests a density of the form: \[\rho_N(x) = \begin{cases}
	\beta_N (\gamma_N - x)^{\alpha_N} & 0 \leq x < \gamma_N \\
	\frac{1}{1-\gamma_N} & \gamma_N \leq x \leq 1
\end{cases}\] for some $\alpha_N$, $\beta_N$ and $\gamma_N$. Thus, by fitting such a $\rho_N$ to a given histogram, we have the estimate $x^* \approx \gamma_N$.
By performing this fit for each trial (at a given population level), we have a set of estimated values of $\alpha_N, \beta_N$ and $\gamma_N$ summarized in Table \ref{scaling_data}.
See also Figure \ref{scaling_parameters}. Note that for this method we pooled all data at a given population leve (from both RNGs) since the previous methods indicated that the
difference between them was minimal and allowed for a greater amount of data.

\begin{figure}[]
    \centerline{\includegraphics[width=350px]{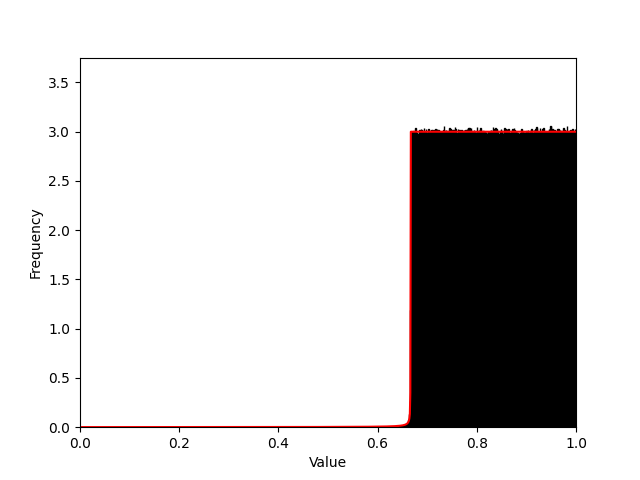}}
    \caption{\textbf{Method 3.} One fitness histogram at $N=25600$ with the best-fit curve $\rho_N$ superimposed.\label{scaling_example}}

    \centerline{\includegraphics[width=350px]{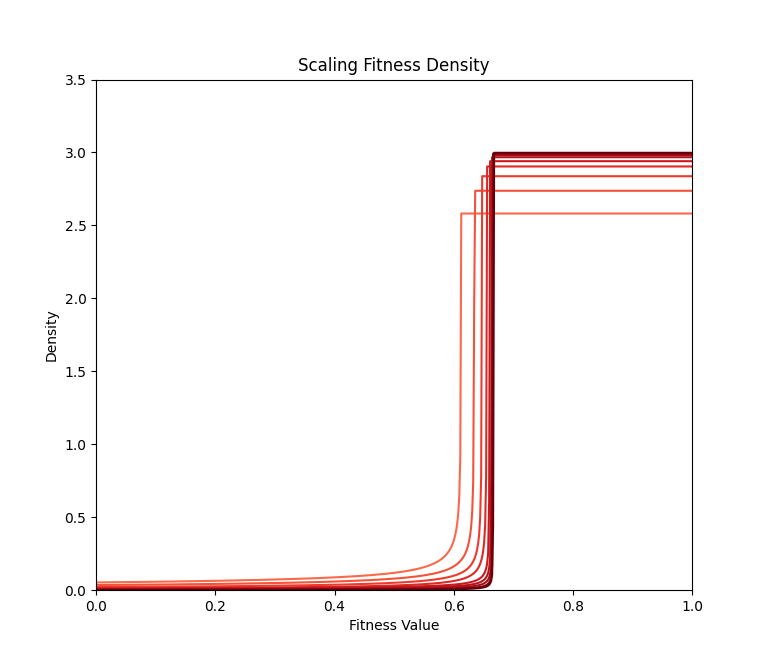}}
    \caption{\textbf{Method 3.} The range of best-fit curves $\rho$ for each population level $N$ are superimposed, exhibiting a clear
approach to a uniform distribution. Note that the maximum value of each density is increasing as $N$ increases.      \label{scaling_density}}

\end{figure}

\begin{figure}[]
    \centerline{\includegraphics[width=350px]{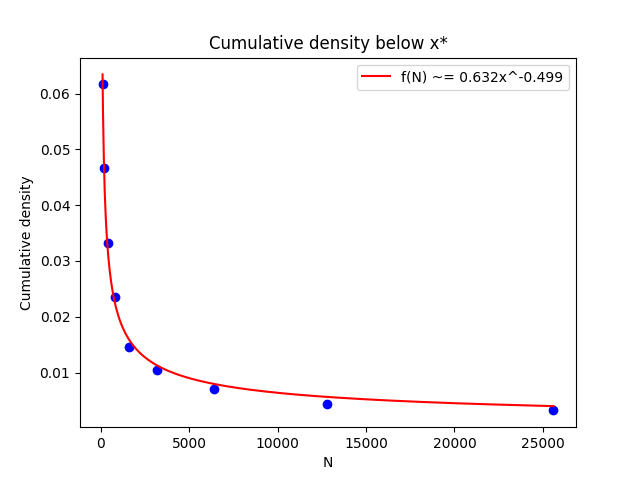}}
    \caption{\textbf{Method 3.} Estimated cumulative density in the interval $(0,x^*)$ for each population level is determined by numerically integrating
the best-fit density curve $\rho_N$. This quantity vanishes as $\rho_N$ approaches a uniform distribution in the limit $N \to \infty$.}
    \label{integral}
\end{figure}

Modeling the fitness histograms with the scaling density $\rho_N$ clearly exhibits the approach to a uniform distribution (see Figure \ref{scaling_density}).
It also allows for a simple estimate of the proportion of the density lying below the cutoff as $N \to \infty$ (see Figure \ref{integral}).

\subsection{\textit{Estimating $x^*$}}

Using the estimates for the values of $x^*_{N}$ determined by each method, we performed a statistical estimation of the cutoff
at infinity $x^* = \lim_{N\to \infty} x^*_{N}$ using maximum likelihood estimation (MLE). Supposing that
the values $x^*_{N}$ follow the function
\begin{equation}
x^*_{100\cdot 2^i}:= f(i) = x^* - \kappa \lambda^i \,,
\label{f(i)}
\end{equation}
we used MLE to estimate the parameters $x^*$, $\kappa$, and $\lambda$. In particular, for $i \in \{0, ..., 8\}$, let $\overline{x}_i$
be the mean obtained via histogram and let $s_i$ be the corresponding standard deviation (see Table \ref{histogramdata}).
Then determine $x^*$, $\kappa$ and $\lambda$ that minimize
\[
\chi^2 (x^*, \kappa, \lambda) = \sum_{n=0}^{8} \frac{(\overline{x}_i - f(i))^2}{{s_i}^2} \,,
\]
where $f(i)$ is given by \eqref{f(i)}. We minimized $\chi^2$ via gradient method with a variable
step\footnote{In particular, we performed the backtracking line search method to determine the step size.}.
This resulted in the estimated values for $x^*$, $\kappa$, and $\lambda$, seen in Table \ref{mle_results},
for each RNG.

\begin{table}[]\centering

    \begin{tabular}{lllll}
    Source & $x^*$ & $\lambda$ & $\kappa$ & $\chi^2$  \\  \cline{1-5}
&&&&\\
     \textbf{Method 1}& & & & \\
	Mersenne & 0.66688&	0.50525&	0.05422&  3.99643\\
	Xoshiro & 0.66697	&0.53743&	0.04442& 13.72710\\
	Combined & 0.66690&	0.51008&	0.05242& 2.62146 \\
&&&&\\
 \textbf{Method 2}&&&&\\
	Mersenne & 0.66717	&0.60361&	0.05547	&0.53577 \\
	Xoshiro & 0.66716&	0.60421	&0.05500&	0.26683\\
	Combined & 0.66717&	0.60357&	0.05563	&0.41982 \\
&&&&\\
 \textbf{Method 3}&&&&\\
	Combined & 0.66748&	0.60659	&0.05441&	0.72618
    \end{tabular}
    \caption{Initial optimized parameters minimizing the $\chi^2$ function, given the data in Table \ref{histogramdata}.
The minimum $\chi^2$ value attained via these parameters is reported. These constitute a single set of optimized parameters,
whereas the randomized data method of Section 4.3 provides a range of values for these parameters (see Table \ref{errors}).}
    \label{mle_results}
\end{table}

We take the $x^*$, $\kappa$, and $\lambda$ associated with the combined data to be our best initial estimates.
Beginning with these values, we determine our final estimates via the randomized data method described below.

\subsection{\textit{Estimating errors on parameters}}

\begin{figure}[]
    \centering
        \begin{tabular}{cc}
            \subfloat[Method 1, Mersenne Twister]{\includegraphics[width=220px]{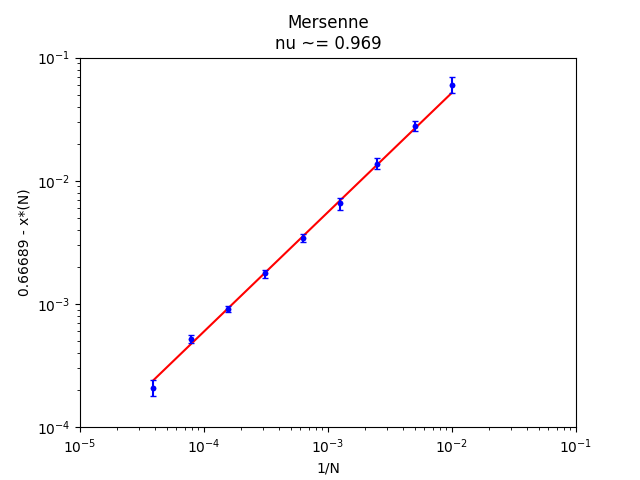}} & \subfloat[Method 2,  Mersenne Twister]{\includegraphics[width=220px]{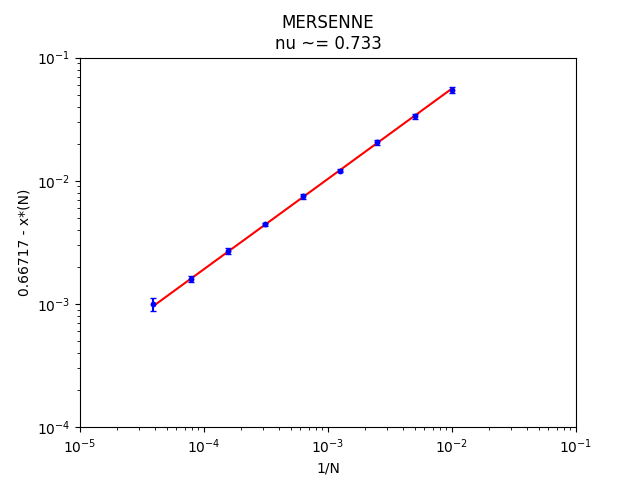}}\\
            \subfloat[Method 1, Xoshiro]{\includegraphics[width=220px]{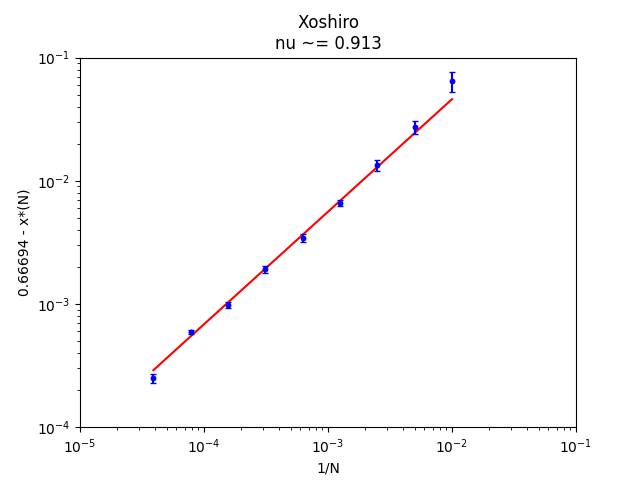}} & \subfloat[Method 2, Xoshiro]{\includegraphics[width=220px]{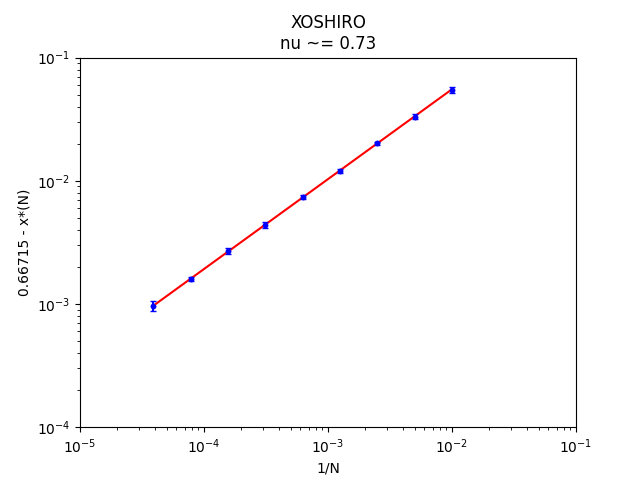}} \\
            \subfloat[Method 1, Combined]{\includegraphics[width=220px]{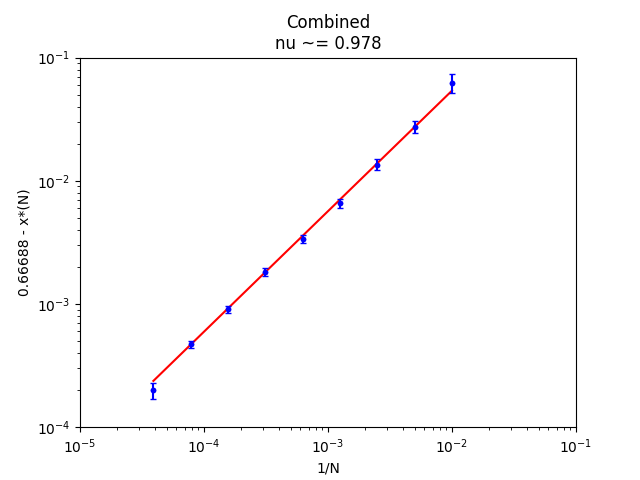}} &\subfloat[Method 2, Combined]{\includegraphics[width=220px]{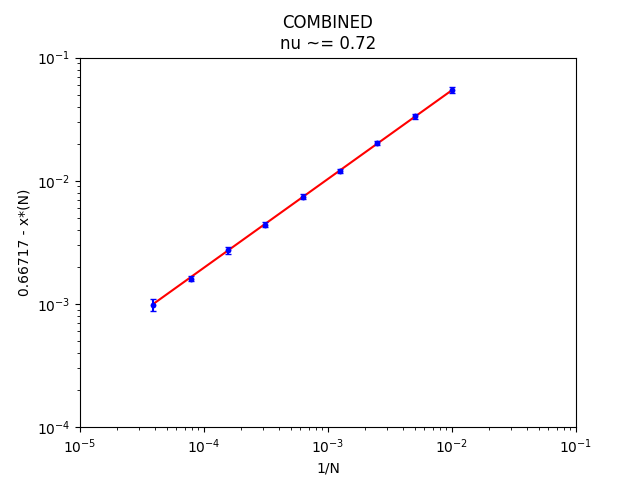}}
        \end{tabular}
        \caption{Reformulating the model $x^*_{100\cdot2^i} = f(i) = x^* - \kappa \lambda^i$ as a function of $\frac{1}{N}$,
        we can write $x^*-x^*_N = \left(\kappa\, \lambda^{-\log_2 100}\right) N^{-\nu}$, where $\nu = -\ln{\lambda}/\ln{2}$.
        The figures show both the original $x^*_N$ data alongside the model, using the mean parameters found in Table \ref{errors}.
        Note that $\nu$ is the slope of the line, and is smaller than previously reported (e.g. \cite{garcia} gives $\nu \approx 1.4$).}
        \label{nu_graphs}
\end{figure}

\begin{figure}[]
    \centering
        \includegraphics[width=240px]{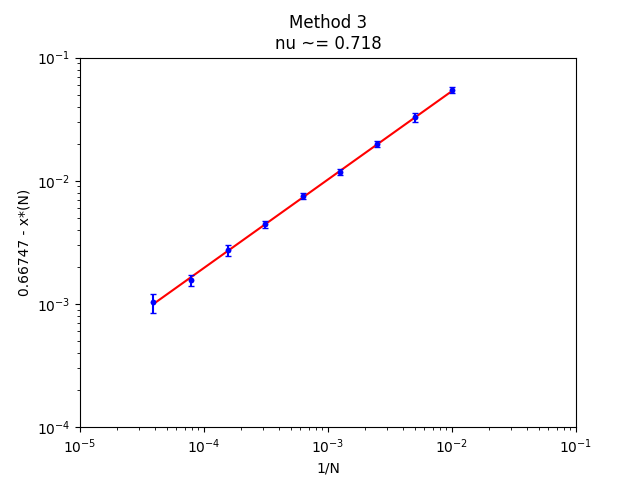}
        \caption{The fit of $x^*-x^*_N = \left(\kappa\, \lambda^{-\log_2 100}\right) N^{-\nu}$ to the cutoff values $x^*_N$  obtained
via Method 3. Note that the value of $\nu$ approximately agrees with the value found in Method 2.}
        \label{nu_graphs2}
\end{figure}

Having obtained an estimate for the cutoff, we estimated the error of our estimate using the following
randomized data method. Initially, for each population level, and each RNG, we obtained an empirical cutoff value
$\overline{x}^*_i$ as a mean over $20$ executions, and also obtained a standard deviation $s_i$.
Supposing that these data were drawn from a normal distribution $\mathcal{N}$ with mean
$\overline{x}^*_i$ and standard deviation $s_i$, we drew $10$ new samples from $\mathcal{N}$ and
performed the same MLE process described above, beginning with these artificial data. This
process provided larger set of possible cutoff values, whose mean and standard deviation we report in
Table \ref{errors}. We take these means to be our best estimates for the Bak-Sneppen cutoff given by each method.
We also report the mean and standard deviation of the associated values of $\lambda$, $\kappa$, and $\nu$.
See below for a discussion of $\nu$, as well as Figures \ref{nu_graphs} and \ref{nu_graphs2} illustrating 
models for $x^* - x^*_N$ fit with these mean parameters.

\begin{table}[]
\centering
    \begin{tabular}{lllllllll}
    Source & Mean $x^*$ & $\pm$ & Mean $\lambda$ & $\pm$ & Mean $\kappa$ &$\pm$ & Mean $\nu$ & $\pm$ \\  \cline{1-9}
&&&&\\
     \textbf{Method 1}& & & & \\
    Mersenne & 0.66689 & 0.00004 &  0.51075 &  0.00631 & 0.05194 & 0.00284 & 0.96930 & 0.01771\\
    Xoshiro & 0.66694 & 0.00004 &  0.53091 & 0.01020 & 0.04597 & 0.00265 & 0.91346 & 0.02744\\
    Combined & 0.66688 & 0.00003 &  0.50758 & 0.00860 & 0.05393 & 0.00291 & 0.97830 & 0.02425\\
&&&&\\
     \textbf{Method 2}& & & & \\
    Mersenne & 0.66717&	0.00014&	0.60166&	0.00770&	0.05609&	0.00172&	0.73298	&0.01834\\
    Xoshiro & 0.66715&	0.00014&	0.60308&	0.00846&	0.05536&	0.00135&	0.72958&	0.02010\\
    Combined & 0.66717&	0.00010&	0.60690	&0.00713&	0.05444	&0.00173	&0.72048&	0.01685\\
&&&&\\
     \textbf{Method 3}& & & & \\
    Combined &0.66747&	0.00025&	0.60799&	0.01142&	0.05355&	0.00176&	0.71788&	0.02685

    \end{tabular}
    \caption{Mean cutoff values and estimated standard deviation obtained via the randomized data method, for each
    RNG. Mean and standard deviation of associated $\lambda$, $\kappa$, and $\nu$ are also reported. To obtain the ``combined''
    results, we pooled together empirical cutoff data from both RNGs before performing the randomization procedure.
 As before, we report the ``Combined'' data for our best estimates.}
    \label{errors}
\end{table}

\subsection{\textit{Estimating the exponent $\nu$}}

To estimate $\nu$ in
\[
x^*-x^*_N\propto N^{-\nu} \,,
\]
we need to rewrite this equation in terms of the parameters used so far. Recall that $x^*_{100\cdot 2^i}$
is written as $f(i)$. So from \eqref{f(i)}, while noting that $i=\log_2(N/100)$, we obtain
\[
x^*-x^*_N = \kappa \,\lambda^i=\kappa\, \lambda^{\log_2N-\log_2 100}=
\left(\kappa\, \lambda^{-\log_2 100}\right) \,\lambda^{\log_2 N} \propto N^{\ln\lambda/\ln 2} \,,
\]
since $\log_2N=\ln N/ \ln 2$. Thus we obtain
\[
\nu=-\ln \lambda/ \ln 2 \,.
\]

To get an idea, take the first two values for $\lambda$ in Table \ref{mle_results}. This gives $\nu=0.985$
and $\nu = 0.896$. Thus $\nu$ is particularly sensitive to variations in $\lambda$. We found an estimated
range of values for $\kappa$ and $\lambda$, and the corresponding range of values for $\nu$, as reported
in Table \ref{errors}.

\section{Conclusion}

We summarize our results in Table \ref{finalvals} (see also Figure \ref{cutoffs}).

\begin{table}[h] \centering
    \begin{tabular}{lllll}
    Source & Estimated $x^*$ & $\pm$ & Minimum & Maximum \\ \cline{1-5}
     &  &  &  &  \\
    Grassberger~\cite{grassberger} & 0.66702 & 0.00008 & 0.66694 & 0.66710 \\
    Paczuski~\cite{pac} & 0.66702 & 0.00004 & 0.66698 & 0.66706 \\
    Garcia~\cite{garcia} & 0.66720 & 0.00020 & 0.66700 & 0.66740 \\
	&&&&\\
    \textbf{Method 1} & 0.66688 & 0.00004 & 0.66684 & 0.66692\\
    \textbf{Method 2} & 0.66717	&0.00010&	0.66707	&0.66726\\
   \textbf{Method 3} &  0.66747& 0.00025 &  0.66722&	0.66772

    \end{tabular}
    \caption{Summary of our simulation results compared to cutoff values previously estimated. The minimum and maximum values (based on the reported error) are listed for ease of comparison.}
\label{finalvals}
\end{table}

Broadly, these results agree with the values previously obtained in the literature. As our simulations did not make use of the scaling assumptions found in e.g. \cite{pac}, this suggests that these assumptions are reasonable and, in particular, that any error they introduce in the cutoff value cannot be much greater than $\approx 0.66740 - 0.66684 = 0.00056$ (the difference between the maximum estimated value under scaling assumptions and the minimum estimated value under direct simulation).

Method 1, as expected, provides an underestimate of the cutoff, illustrated most dramatically by Figure \ref{nu_graphs},
in which the underestimate at the lowest population values skews $\nu$, the key parameter on the overall fit of these data
to a curve $x^*-x^*_N\propto N^{-\nu}$. This, along with the fact that it is significantly lower than reported in the literature
indicates that Method 1 is not ideal for measuring the cutoff, and that the value of $\nu$ it provides should not be given much
weight. However, Method 1 did provide the narrowest range of possible values, suggesting that an appropriately modified
version could provide the most accurate estimates of the true cutoff value.

The estimates provided by Methods 2 and 3, while less precise than Method 1, do generally agree with the literature
and take into account more information about the shape of the distribution. Method 3 in particular closely models the
density on both sides of the cutoff and suggests a deeper exploration of the scaling properties of the fitness distribution.
Both methods give a similar value for $\nu \approx 0.72$, which is notably lower than previously measured (see \cite{garcia}).
The cutoff value given by Method 2 is lower than the value given by Method 3, likely because of its discussed tendency to
underestimate the height of the true distribution (and hence the value $x^* = 1 - h^{-1}$).

Taking the value $x^* = 0.66747 \pm 0.00025$ obtained by Method 3 to be our best estimate, we note that it is significantly larger that other values reported in the literature. This suggests that a deeper investigation of the scaling law approach
is warranted in order to more precisely determine the true value of $x^*$.

\section{Data availability statement}

The data that support the findings of this study are available from the corresponding author, Cameron
Fish, upon request. The simulation is written in C and the subsequent statistical analysis
was performed with Python. The code for both can be found at \texttt{https://github.com/camafish/Bak-Sneppen}.

\vskip 0.1in\noindent
{\bf Acknowledgement.} We are grateful to Javier Prieto and Daniel Taylor-Rodriguez for their generous
help with the statistics involved in this work.

\bibliography{researchbib}
\bibliographystyle{plain}

\end{document}